\documentclass[%
 reprint,
showpacs,
 amsmath,amssymb,
 aps,
floatfix
]{revtex4-1}

\usepackage{graphicx}
\usepackage{dcolumn}
\usepackage{bm}
\usepackage{float}
\usepackage{flafter}
\restylefloat{table}
\usepackage{graphicx}
\usepackage{hyperref}
\usepackage{nicefrac}
\usepackage[dvipsnames]{xcolor}
\usepackage{multirow}
\usepackage[normalem]{ulem} 
\newcommand{\stkout}[1]{\ifmmode\text{\sout{\ensuremath{#1}}}\else\sout{#1}\fi}

\begin{document}


\title{Asymptotically Unbiased Estimation of Physical Observables with Neural Samplers}

\author{Kim A. Nicoli}
\affiliation{Machine Learning Group, Technische Universit\"{a}t Berlin, 10587 Berlin, Germany}

\author{Shinichi Nakajima}
\affiliation{Machine Learning Group, Technische Universit\"{a}t Berlin, 10587 Berlin, Germany}
\affiliation{Berlin Big Data Center, 10587 Berlin, Germany}
\affiliation{RIKEN Center for AIP, 103-0027, Tokyo, Japan}

\author{Nils Strodthoff}
\affiliation{Fraunhofer Heinrich Hertz Institute, 10587 Berlin, Germany}

\author{Wojciech Samek}
\email{wojciech.samek@hhi.fraunhofer.de}
\affiliation{Fraunhofer Heinrich Hertz Institute, 10587 Berlin, Germany}
\affiliation{Berlin Big Data Center, 10587 Berlin, Germany}
\affiliation{Berliner Zentrum f\"ur Maschinelles Lernen, 10587 Berlin, Germany}

\author{Klaus-Robert M\"{u}ller}
\email{klaus-robert.mueller@tu-berlin.de}
\affiliation{Machine Learning Group, Technische Universit\"{a}t Berlin, 10587 Berlin, Germany}
\affiliation{Berlin Big Data Center, 10587 Berlin, Germany}
\affiliation{Department of Brain and Cognitive Engineering, Korea University, Anam-dong, Seongbuk-gu, Seoul 136-713, South Korea}
\affiliation{Max-Planck-Institut f\"{u}r Informatik, Saarbr\"{u}cken, Germany}
\affiliation{Berliner Zentrum f\"ur Maschinelles Lernen, 10587 Berlin, Germany}

\author{Pan Kessel}
\affiliation{Machine Learning Group, Technische Universit\"{a}t Berlin, 10587 Berlin, Germany}
\affiliation{Berlin Big Data Center, 10587 Berlin, Germany}

\date{\today}

\begin{abstract}
We propose a general framework for the estimation of observables with generative neural samplers focusing on modern deep generative neural networks that provide an exact sampling probability. In this framework, we present asymptotically unbiased estimators for generic observables, including those that explicitly depend on the partition function such as free energy or entropy, and derive corresponding variance estimators. We demonstrate their practical applicability by numerical experiments for the 2d Ising model which highlight the superiority over existing methods. Our approach greatly enhances the applicability of generative neural samplers to real-world physical systems.
\end{abstract}

\maketitle

\section{\label{sec:Introduction}Introduction}
Monte-Carlo methods are the workhorses of statistical physics and lattice field theories providing insights in strongly correlated systems from first principles \cite{gattringerlang, newman1999monte}. In spite of their overall success, these approaches come with a number of downsides: Monte-Carlo methods potentially get trapped in local minima that prevent them from exploring the full configuration space \cite{Kirkpatrick1983:Optimization}. Furthermore, they can suffer from large autocorrelation times --- in particular close to criticality thus making them very costly in certain regions of the parameter space. In these regions, observables at physical parameter values can often only be extrapolated from simulations at unphysical parameter values. Last but not least, observables that explicitly depend on the partition function, such as free energy and entropy, can only be evaluated up to an overall constant by "chaining" the results of a considerable number of Monte-Carlo chains \cite{newman1999monte, bishop2006pattern,nakajima2019vblt}.  

Generative neural samplers (GNSs) are machine learning models which allow sampling from probability distributions learned by using deep neural networks. We refer to \cite{goodfellow2016deep} for an accessible overview. GNSs have shown remarkable performance in generating realistic samples capturing complicated probability distributions of real-world data such as images, speech, and text documents. This has inspired the application of GNSs in the context of theoretical physics \cite{Torlai2016LearningTW,Morningstar2017DeepLT,liu2017simulating,huang2017accelerated, li2018neural,koch2018mutual,Urban:2018tqv,zhou2019regressive,mustafa2019cosmogan,nicoli2019comment,hu2019machine,yang2019deep,albergo2019flow,wu2019solving,sharir2019deep,noe2019boltzmann}.

In this work, we focus on a particularly promising subclass of GNSs. Namely, we will consider deep neural networks $q$ that allow to sample configurations $s \sim q$ from the model and also provide the exact probability $q(s)$ of the sample $s$. A notable example for this type of GNS are Variational Autoregressive Networks (VANs)\cite{wu2019solving}, which sample from a PixelCNN \cite{vanOord2016pixelrnn} to estimate observables. The main advantage of this class of GNSs is that they can be trained without resorting to Monte-Carlo configurations by minimizing the Kullback--Leibler divergence between the model $q$ and a target (Boltzmann) distribution $p$. As a result, they represent a truly complementary approach to existing Monte-Carlo methods. 

Observables are often estimated by directly sampling from the GNS and then taking the sample mean. However, as we will discuss in detail, this approach suffers from a mismatch of the sampling distribution $q$ and the target distribution $p$. This mismatch is unavoidable since it cannot be expected that the GNS fully captures the underlying physics. This leads to uncontrolled estimates as both the magnitude and the direction of this bias is in general unknown and scales unfavorably with the system size \cite{nicoli2019comment}. 

In this work, we propose a general framework to avoid this serious problem. Our method applies to any GNS with exact sampling probability. Specifically, we will show that it is possible to define asymptotically unbiased estimators for observables along with their corresponding variance estimators. Notably, our method also allows to directly estimate observables that explicitly depend on the partition function, e.g. entropy and free energy. Our proposal therefore greatly enhances the applicability of GNSs to real-world systems. 

The paper is organized as follows: In Section~\ref{sec:Background}, we will discuss the proposed asymptotically unbiased estimators for observables along with corresponding variance estimators. We illustrate the practical applicability of our approach for the two-dimensional Ising model in Section~\ref{sec:Experiments}, discuss the applicabilty to other GNSs in Section~\ref{sec:applic} and conclude in Section~\ref{sec:conclusion}. Technical details are presented in several appendices.

\section{\label{sec:Background}Asymptotically Unbiased Estimators}

\begin{figure*}[ht]
\centering
\includegraphics[width=\linewidth]{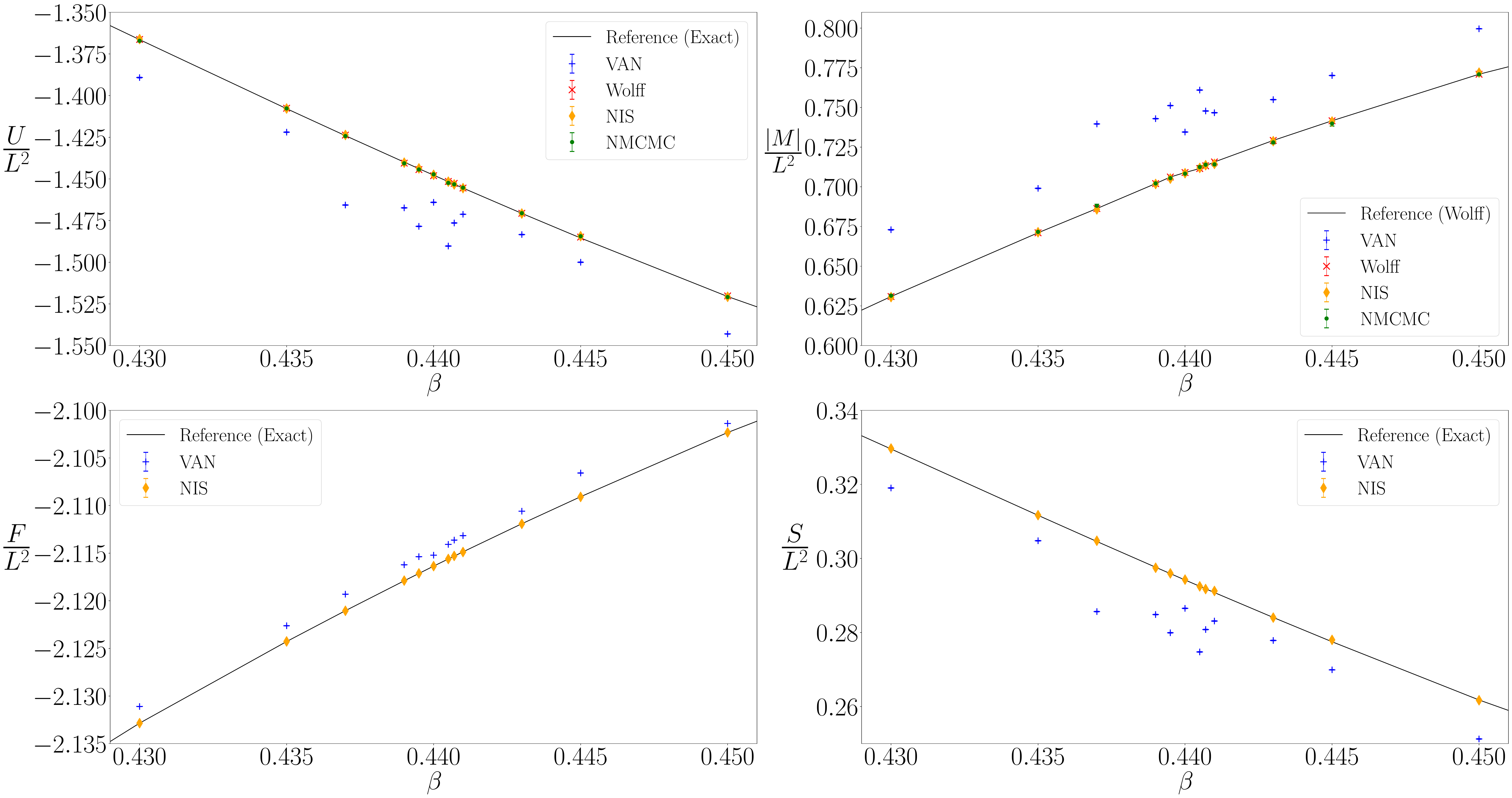}
  \caption{Estimates for various observables around $\beta_c$. NMCMC and NIS agree with the reference values provided by the exact analytical solutions as well as the Wolff algorithm. VAN deviates significantly. Observables are: internal energy $U=\mathbb{E}_p[H]$, absolute magnetization $|M|=\sum_i\mathbb{E}_p(|s_i|)$, the free energy $F=\tfrac{-1}{\beta} \ln(Z)$ and the entropy $S=-\mathbb{E}_p[\ln p]$.}\label{fig:observables}
\end{figure*}

\subsection{\label{sec:VAN}Generative Neural Samplers with Exact Probability (GNSEP)}
We will use a particular subclass of GNSs to model the variational distribution $q$ as they can provide the exact probability $q(s)$ of configurations $s$ and also allow sampling from this distribution $s \sim q$. We will henceforth refer to this subclass as generative neural samplers with exact probability (GNSEP).  Using these two properties, one can then minimize the inverse Kullback--Leibler divergence between the Boltzmann distribution $p(s)=1/Z \, \exp(-\beta H(s))$ and the variational distribution $q$ without relying on Monte-Carlo configurations for training,
\begin{align}
    \textrm{KL} (q | p) &= \sum_s q(s) \, \ln \left(\frac{q(s)}{p(s)}\right) \nonumber \\ &= \sum_s q(s) (\ln(q(s)) + \beta H(s)) + \ln  Z \,. \label{eq:loss}
\end{align}

This objective can straightforwardly be optimized using gradient decent since the last summand is an irrelevant constant shift. After the optimization is completed, observables (expectation values of an operator $\mathcal{O}$ with respect to the Boltzmann distribution $p$) are then conventionally estimated by the sample mean  
\begin{align}
\langle \mathcal{O}(s) \rangle_{p} \approx \frac{1}{N}\sum_{i=1}^N \mathcal{O}(s_i) \label{eq:samplemean}
\end{align}
using the neural sampler $s_i \sim q $. 

Various architectures for generative neural samplers are available. Here, we will briefly review the two most popular ones: 

\paragraph{Normalizing Flows (NFs):} Samples from a prior distribution $q_0(z)$, such as a standard normal, are processed by an invertible neural network $f(z)$. The probability of a sample $s=f(z)$ is then given by
\begin{align*}
    q(s) = q_0 (f^{-1}(s)) \left|\det \left(\frac{\partial f}{\partial z}\right)\right|^{-1} \,.
\end{align*}
The architecture of $f$ is chosen such that the inverse and its Jacobian can easily be computed. Notable examples of normalizing flows include NICE\cite{dinh2014nice}, RealNVP\cite{dinh2016density} and GLOW\cite{kingma2018glow}. First physics applications of this framework have been presented in \cite{noe2019boltzmann} in the context of quantum chemistry and subsequently in \cite{albergo2019flow} for lattice field theory.

\paragraph{Autoregressive Models (AMs):} In this case, an ordering $s_1, \dots, s_N$ of the components of $s$ is chosen and the conditional distribution $q(s_i| s_{i-1} \dots s_1)$ is modeled by a neural network. The joint probability $q(s)$ is then obtained by multiplying the conditionals
\begin{align}
    q(s) = \prod_{i=1}^N q(s_i| s_{i-1} \dots s_1) 
\end{align}
and one can draw samples from $q$ by autoregressive sampling from the conditionals. State-of-the-art architectures often use convolutional neural networks (with masked filters to ensure that the conditionals only depend on the previous elements in the ordering). Such convolutional architectures were first proposed in the context of image generation with PixelCNN \cite{vanOord2016pixelrnn,Salimans2017pixelcnn} as most prominent example. In \cite{wu2019solving} these methods were first used for statistical physics applications.

A major drawback of using generative neural samplers is that their estimates are A) (often substantially) biased and B) do not come with reliable error estimates, see Figure~\ref{fig:observables}. Both properties are obviously highly undesirable for physics applications.
The main reason for this is that the mean \eqref{eq:samplemean} is taken over samples drawn from the sampler $q$ to estimate expectation values with respect to the Boltzmann distribution $p$. However, it cannot be expected that the sampler $q$ perfectly reproduces the target distribution $p$. This discrepancy will therefore necessarily result in a systematic error which is often substantial. Furthermore, in all the cases that we are aware of, this error cannot be reliably estimated.

In order to avoid this serious problem, we propose to use either importance sampling or Markov chain Monte Carlo (MCMC) rejection sampling to obtain asymptotically unbiased estimators. We also derive expressions for the variances of our estimators.

\subsection{Sampling Methods}\label{sec:methods}
Here we propose two novel estimators that are asymptotically unbiased and are shown to alleviate the serious issues A) and B) mentioned in the previous section.\\
\emph{Neural MCMC (NMCMC)} uses the sampler $q$ as the proposal distribution $p_0(s|s')$ for a Markov Chain. Samples $s\sim p_0(s|s')$ are then accepted with probability
\begin{align}
\text{min}\left(1, \tfrac{p_0(s| s') \, p(s')}{p_0(s'| s) \, p(s)}\right)
    = \text{min}\left(1, \tfrac{q(s) \, \exp(-\beta H(s'))}{q(s') \, \exp(-\beta H(s))}\right) \,.
\end{align}
We note that the proposal configurations do not depend on the previous elements in the chain. This has two important consequences: Firstly, they can efficiently be sampled in parallel. Secondly, the estimates will typically have very small autocorrelation. 

\emph{Neural Importance Sampling (NIS)} provides an estimator by
\begin{align}
    \langle \mathcal{O}(s) \rangle_{p} \approx 
    \textstyle \sum_{i} w_i \, \mathcal{O}(s_i) &&\text{with} &&  s_i \sim q  \,,
\end{align}
where $w_i = \tfrac{\hat{w}_i}{\sum_i \hat{w}_i}$ for $\hat{w}_i = \tfrac{e^{-\beta H(s_i)}}{q(s_i)}$ is the importance weight. It is important to stress that we can obtain the configurations $s_i$ by independent identically distributed (iid) sampling from $q$. This is in stark contrast to related reweighting techniques in the context of MCMC sampling \cite{gattringerlang}.

We assume that the output probabilities of the neural sampler $q$ are bounded within $[\epsilon, 1 - \epsilon]$ for small $\epsilon >0$. In practice, this can easily be ensured by rescaling and shifting the output probability of the model as explained in Appendix~\ref{app:proofs}.

It then follows from standard textbook arguments that these two sampling methods provide asymptotically unbiased estimators. For convenience, we briefly recall these arguments in Appendix~\ref{app:proofs}.

We note that our asymptotic unbiased sampling methods have the interesting positive side effect that they allow for \emph{transfer across parameter space}, a property they share with conventional MCMC approaches \cite{NewmanBarkemachap8:1999}. For example, we can use a neural sampler trained at inverse temperature $\beta'$ to estimate physical observable at a different target temperature $\beta\neq \beta'$, as shown later in Section~\ref{sec:Experiments}. In some cases, this can result in a significant reduction of runtime, as we will demonstrate in Section~\ref{sec:Experiments}.

\subsection{Asymptotically Unbiased Estimators}
\begin{figure}[ht]
\centering
\includegraphics[width=\linewidth]{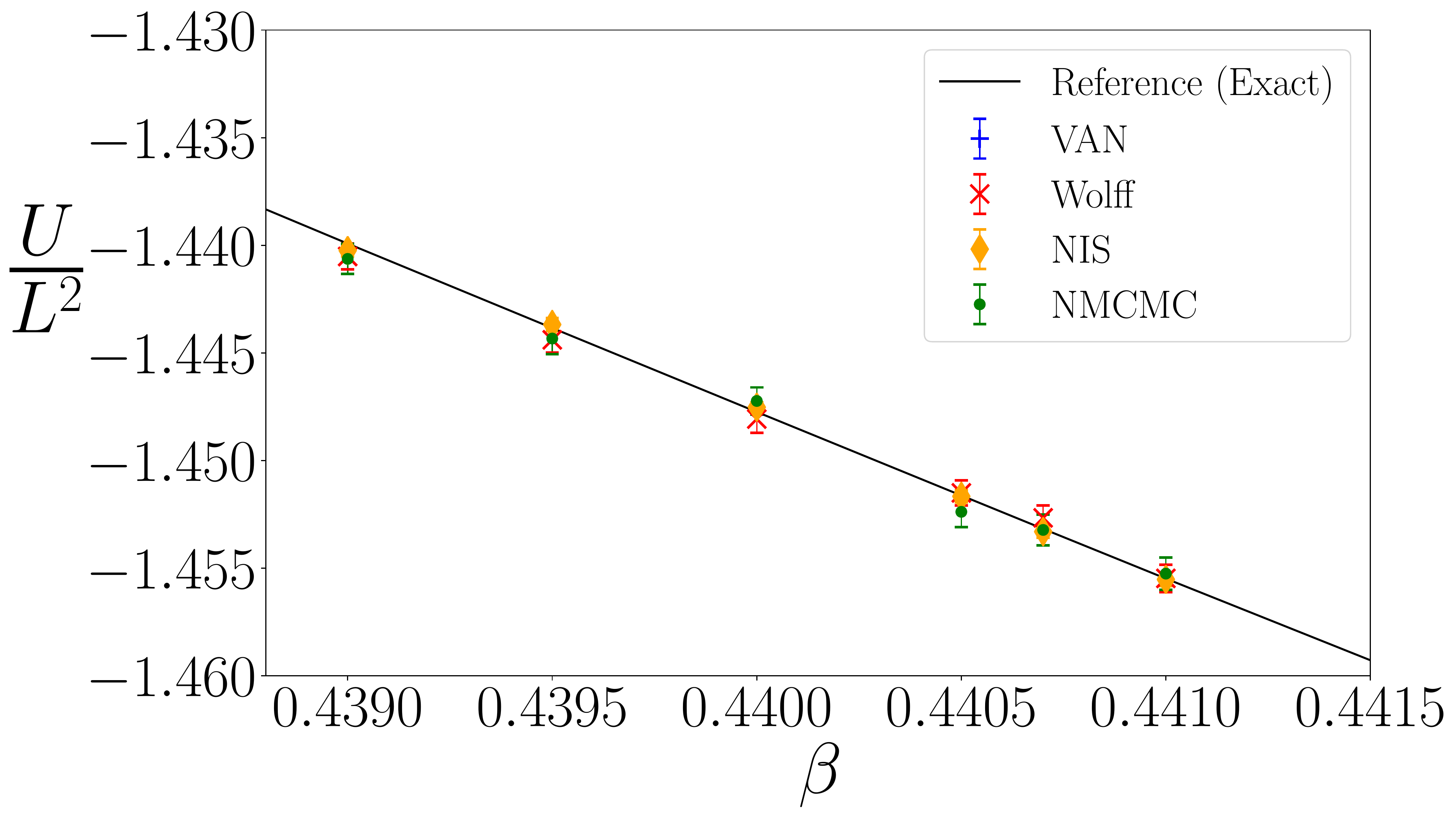}
  \caption{Zoom around critical $\beta_c\approx0.4407$ showing the internal energy per lattice site from Figure~\ref{fig:observables} (e.g. $16 \times 16$ lattice). We took internal energy as a reference example, same considerations hold for the other observables. The estimates from our proposed method match with the exact reference. VAN is not shown because it is out of range as one can see from Figure~\ref{fig:observables}.}\label{fig:zoom_energy}
\end{figure}
For operators $\mathcal{O}(s)$ which do not explicitly depend on the partition function, such as internal energy $\mathcal{O}_U(s) = H(s)$ or absolute magnetization $\mathcal{O}_{|M|}(s) = \sum_i |s_i|$, both NIS and NMCMC provide asymptotically unbiased estimators as explained in the last section.

However, generative neural samplers are often also used for operators $\mathcal{O}(s, Z)$ explicitly involving the partition function $Z$. Examples for such quantities include
\begin{align}
    &\mathcal{O}_F(s,Z) = -\frac{1}{\beta} \ln(Z)\,,\label{eq:obsF} \\ &\mathcal{O}_S(s,Z) = \beta \, H(s) + \ln Z \,, \label{eq:obsS}
\end{align}
which can be used to estimate the free energy $F=\tfrac{-1}{\beta} \ln(Z)=\mathbb{E}_p[ \mathcal{O}_F ]$ and the entropy $S=-\mathbb{E}_p[\ln p]=\mathbb{E}_p[ \mathcal{O}_S ]$ respectively. Since the Kullback-Leibler divergence is greater or equal to zero, it follows from the optimization objective \eqref{eq:loss} that 
\begin{align}
  F_q = \frac{1}{\beta} \sum_s q(s) (\ln(q(s)) + \beta H(s)) \ge - \frac{1}{\beta} \ln(Z) = F \,.
\end{align}
Therefore, the variational free energy $F_q$ provides an upper bound on the free energy $F$ and is thus often used as its estimate. Similarly, one frequently estimates the entropy $S=-\mathbb{E}_p(\ln p)$ by simply using the variational distribution $q$ instead of $p$. Both estimators however typically come with substantial biases which are hard to quantify. This effect gets particularly pronounced close to the critical temperature.

Crucially, neural importance sampling also provides asymptotically unbiased estimators for $\mathbb{E}_p[\mathcal{O}(s,Z)]$ by
\begin{align}
    \hat{\mathcal{O}}_N = \frac{\tfrac{1}{N}\sum_{i=1}^N \mathcal{O}(s_i, \hat{Z}_N) \, \hat{w}(s_i)}{\hat{Z}_N} && \text{with} && s_i \sim q \,, \label{eq:estOp}
\end{align}
where the partition function $Z$ is estimated by
\begin{align}
   \hat{Z}_N = \frac{1}{N} \, \sum_{i=1}^N \hat{w}_i \,. \label{eq:estimatorZ}
\end{align}
In the next section, we will derive the variances of these estimators. Using these results, the errors of such observables can systematically be assessed. 

\subsection{\label{sec:VarianceEstimators}Variance Estimators}
In the following, we focus on observables of the form
\begin{align}
    \mathcal{O}(s,Z) = g(s) + h(Z) \,,
    \label{eq:GeneeralObservable}
\end{align}
that include the estimators for internal energy, magnetization but most notably also for free energy \eqref{eq:obsF} and entropy \eqref{eq:obsS}.
As just mentioned, expectation values of these operators can be estimated using \eqref{eq:estOp}.

Let us assume that $h$ is differentiable at the true value of the partition function $Z$.
Then, as shown in Appendix~\ref{app:variance},
the variance of the estimator for large $N$ is given by
\begin{align}
    \sigma^2_{\hat{\mathcal{O}}_N} 
=& 
\frac{\boldsymbol{\psi}^T \mathbb{E}_q [\boldsymbol{\phi} \boldsymbol{\phi}^T] \boldsymbol{\psi}}{N} 
+  o_P(N^{-1}) \,, \label{eq:fullvariance}
\end{align}
where
\begin{align}
\boldsymbol{\phi} &=
\begin{pmatrix}
g \hat{w} - \mathbb{E}_p[g] Z\\
\hat{w} - Z
\end{pmatrix},
&
\boldsymbol{\psi} &= 
\begin{pmatrix}
1/Z\\
-  \mathbb{E}_p[g]/ Z + h'(Z) 
\end{pmatrix}.
\end{align}
Note that
$\mathbb{E}_p[g]$ can be estimated by
\begin{align}
\frac{   \frac{1}{N} \sum_{i=1}^N g(s_i) \hat{w}(s_i) }{\hat{Z}_N}
\end{align}
and $Z$ can be estimated by \eqref{eq:estimatorZ}, respectively.

For operators with $h\equiv 0$, it is well-known \cite{isvariance} that Eq.~\eqref{eq:fullvariance} reduces to 
\begin{align}
\sigma^2_{\hat{\mathcal{O}}_N}
&= 
\frac{\text{Var}_p(g)}{N_{\textrm{eff}}} + o_P(N^{-1}) \,,
\end{align}
where we have defined the effective sampling size 
\begin{align}
N_{\textrm{eff}} = \frac{N}{\mathbb{E}_q \left[w^2\right]} \,.
\end{align}
Note that the effective sampling size does not depend on the particular form of $g$. It is however important to stress that for observables with $h\neq 0$, the error cannot be estimated in terms of effective sampling size but one has to use \eqref{eq:fullvariance}. While this expression is more lengthy, it can be easily estimated. Therefore, neural importance sampling allows us to reliably estimate the variances of physical observables --- in particular observables with explicit dependence on the partition function. This is in stark contrast to the usual GNS approach.

It is also worth stressing that MCMC sampling does not allow to directly estimate those observables which explicitly involve the partition function. For completeness, we also note that a similar well-known effective sampling size can be defined for MCMC
\begin{align}
N_{\textrm{eff}} = \frac{N}{2 \, \tau_{int, \mathcal{O}}} \,, \label{eq:autocor}
\end{align}
where $\tau_{int, \mathcal{O}}$ is the integrated auto-correlation time of the operator $\mathcal{O}$, see \cite{wolff2004monte, gattringerlang} for more details.

\section{\label{sec:Experiments}Numerical Results}

We will now demonstrate the effectiveness of our method on the example of the two-dimensional Ising model with vanishing external magnetic field. This model has an exact solution and therefore provides us with a ground truth to compare to. The Hamiltonian of the Ising model is given by
\begin{equation}
    H(s) = -J \sum_{\langle i,j \rangle} s_i \, s_j \,,
\end{equation}
where $J$ is the coupling constant and the sum runs over all neighbouring pairs of lattice sites. The corresponding Boltzmann distribution is then given by
\begin{align}
    p(s) = \frac{1}{Z} \exp(-\beta H(s)) \,,
\end{align}
with partition function $Z=\sum_s \exp(-\beta H(s))$. For simplicity, we will absorb the coupling constant $J$ in $\beta$ in the following. Here, we will only consider the ferromagnetic case for which $J>0$ and the model undergoes a second-order phase transition at $\beta_c\approx0.4407$ in the infinite volume limit.

In addition to the exact solution by Onsager for the infinite volume case \cite{onsager}, there also exists an analytical solution for finite lattices \cite{ferdinand1969bounded}, which we review in Appendix~\ref{app:ZIsing} and use for reference values. An exact partition function for the case of vanishing magnetic field is not enough to derive expressions for some observables, such as magnetization. For these observables, we obtain reference values by using the Wolff MCMC clustering algorithm \cite{wolff1989comparison}. 

\subsection{Unbiased Estimators for the Ising Model}
For discrete sampling spaces, autoregressive algorithms are the preferred choice as normalizing flows are designed for continuous ones \footnote{However, \cite{tran2019discrete, hoogeboom2019integer} present a recent attempt to apply normalizing flows to discrete sampling spaces.}.  It is nonetheless important to stress that our proposed method applies irrespective of the particular choice for the sampler.

We use the standard VAN architecture for the GNS. For training, we closely follow the procedure described in the original publication \cite{wu2019solving}. More detailed information about hyperparameter choices can be found in Appendix~\ref{appendix:hyperparams}. We use VANs, trained for a $16 \times 16$ lattice at various temperatures around the critical point, to estimate a number of observables. The errors for neural importance sampling are determined as explained in Section~\ref{sec:VarianceEstimators}. For Wolff and Neural MCMC, we estimate the autocorrelation time as described in \cite{wolff2004monte}.

Figure~\ref{fig:observables} summarizes the main results of our experiments in terms of estimates for internal energy, absolute magnetization, entropy and free energy around the critical regime. NMCMC and NIS agree with the reference values while VAN deviates significantly. We note that this effect is also present for observables with explicit dependence on the partition function, i.e. for entropy and free energy. 

All estimates in Figure~\ref{fig:observables} deviate from the reference value in the same direction. Whereas this is expected for the free energy (for which the true value is a lower bound) also for the other observables the trained GNSs seem to favor a certain direction of approaching the true value. However, as we show in Appendix~\ref{app:directionofbias}, this trend holds only on average and is not a systematic effect.

\begin{figure}[ht]
    \centering
    \includegraphics[width=1\linewidth]{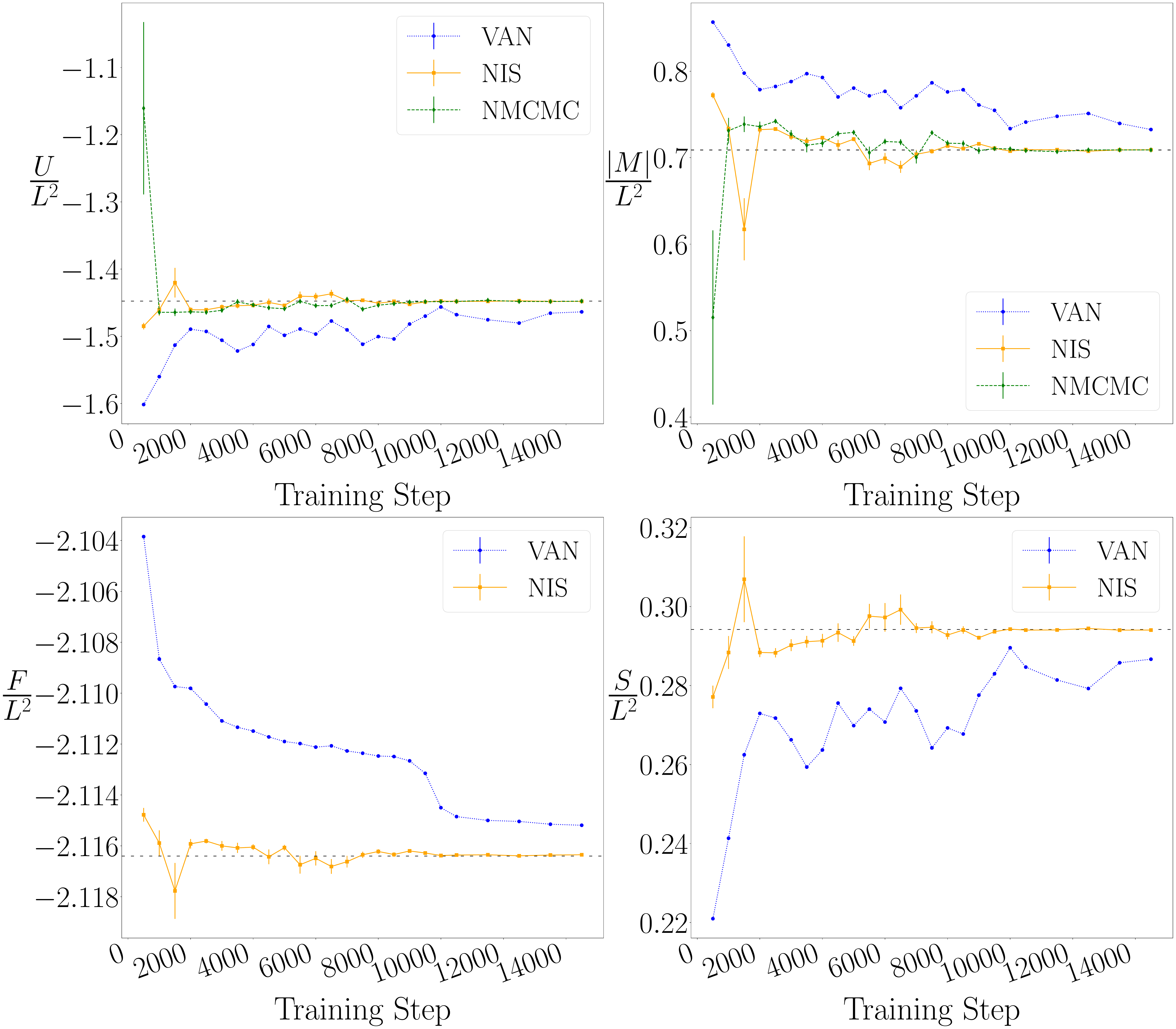}
    \caption{\label{fig:checkpoints} Estimation of observables during training for a single run on a $16\times16$ lattice. The modified sampling procedure leads to accurate predictions at significantly earlier stages of training since it correct for imperfect samplers. As before, we look at the internal energy $U=\mathbb{E}_p[H]$, the absolute magnetization $|M|=\sum_i\mathbb{E}_p(|s_i|)$, the free energy $F=\tfrac{-1}{\beta} \ln(Z)$ and the entropy $S=-\mathbb{E}_p[\ln p]$.}
\end{figure}

In Figure~\ref{fig:checkpoints}, we track the evolution of the estimates for the four observables under consideration during training. 
This figure clearly demonstrates that our proposed method leads to accurate predictions even at earlier stages of the training process. This is particularly important because the overall runtime for GNS estimates is heavily dominated by the training. 

Table~\ref{tab:scaling} summarizes results for $24 \times 24$ lattice. For this larger lattice, the systematic error of VAN is even more pronounced and the estimated values do not even qualitatively agree with the reference values. Our modified sampling techniques, on the other hand, lead to fully compatible results.

\begin{table*}[ht]
    \caption{Comparison of VAN, NMCMC and NIS on a $24\times24$ and a $16\times16$ lattices, both trained at $\beta_c$. Entropy and free energy cannot be directly estimated using Monte Carlo approaches. Bold numbers denote estimates which are compatible with ground truth within one standard deviation. Standard deviations are in parentheses.Observables are: internal energy $U=\mathbb{E}_p[H]$, absolute magnetization $|M|=\sum_i\mathbb{E}_p(|s_i|)$, the free energy $F=\tfrac{-1}{\beta} \ln(Z)$ and the entropy $S=-\mathbb{E}_p[\ln p]$. Details on the runtime performance are reported in Appendix \ref{appendix:hyperparams}.}
    \label{tab:scaling}
    \centering
    \begin{tabular}{c|l|cccc} 
    Lattice & Sampler & $\nicefrac{U}{L^2}$ & $\nicefrac{|M|}{L^2}$ & $\nicefrac{S}{L^2}$ & $\nicefrac{F}{L^2}$ \\
    \colrule
    \multirow{ 3}{*}{(24x24)} & \textbf{VAN}  & -1.5058 (0.0001) & 0.7829 (0.0001) & 0.26505 (0.00004) & -2.107250 (0.000001) \\
    & \textbf{NIS}  & \textbf{-1.43} (0.02) & \textbf{0.67} (0.03) & \textbf{0.299}     (0.007) & \textbf{-2.1128} (0.0008) \\
    & \textbf{NMCMC}  & \textbf{-1.448} (0.007) & \textbf{0.68} (0.04) & - & -\\
    \colrule
    \colrule
    \textbf{Reference} & & -1.44025 & 0.6777 (0.0006) & 0.29611 & -2.11215 \\ 
    \colrule
    \multirow{ 3}{*}{(16x16)} & \textbf{VAN}  & -1.4764 (0.0002) & 0.7478 (0.0002) & 0.28081 (0.00007) & -2.11363 (0.00001) \\
    & \textbf{NIS}  & \textbf{-1.4533} (0.0003) & \textbf{0.71363} (0.00004) & \textbf{0.2917}  (0.0002) & \textbf{-2.11529} (0.00001) \\
    & \textbf{NMCMC}  & \textbf{-1.4532} (0.0007) & \textbf{0.714} (0.001) & - & -\\
    \colrule
    \colrule
    \textbf{Reference} & & -1.4532 & 0.7133 (0.0008) & 0.29181 & -2.11531
    \end{tabular}
\end{table*}

\begin{figure}[ht]
    \centering
    \includegraphics[width=1.\linewidth]{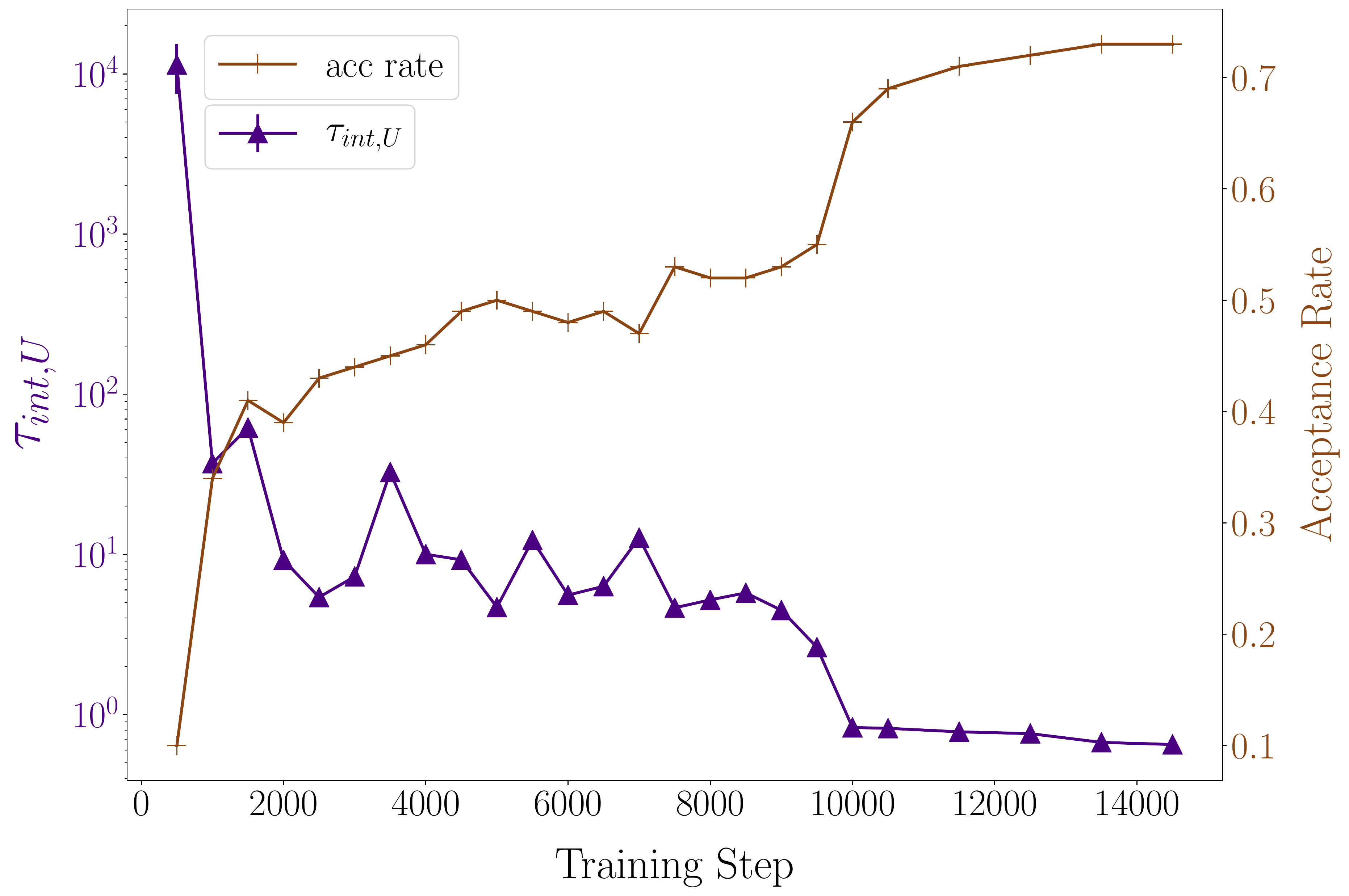}
    \caption{\label{fig:autocorr} Evolution of the acceptance rate (right) and the integrated autocorrelation time of the internal energy $\tau_{int, U}$ (left) during training. NMCMC runs were preformed on a 16$\times$16 lattice at $\beta_c$.}
\end{figure}

\begin{figure}[ht]
    \centering
    \includegraphics[width=1.\linewidth]{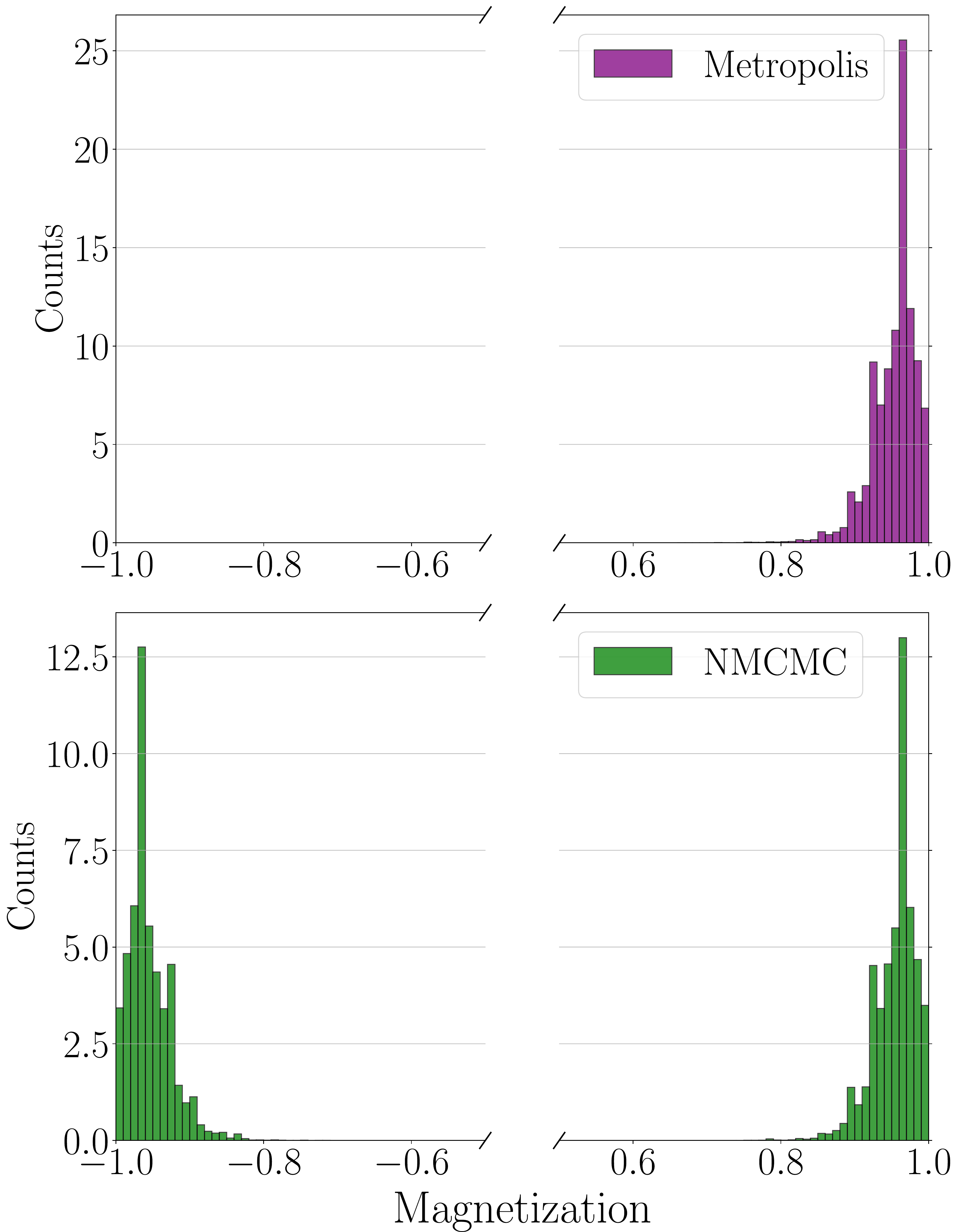}
    \caption{Histogram for the magnetization of the system at $\beta=0.55$. While the Metropolis algorithm is only able to capture one of the two modes of the distribution, NMCMC is able to cover both.}\label{fig:histogram}
\end{figure}

\begin{figure}[ht]
    \centering
    \includegraphics[width=1.\linewidth]{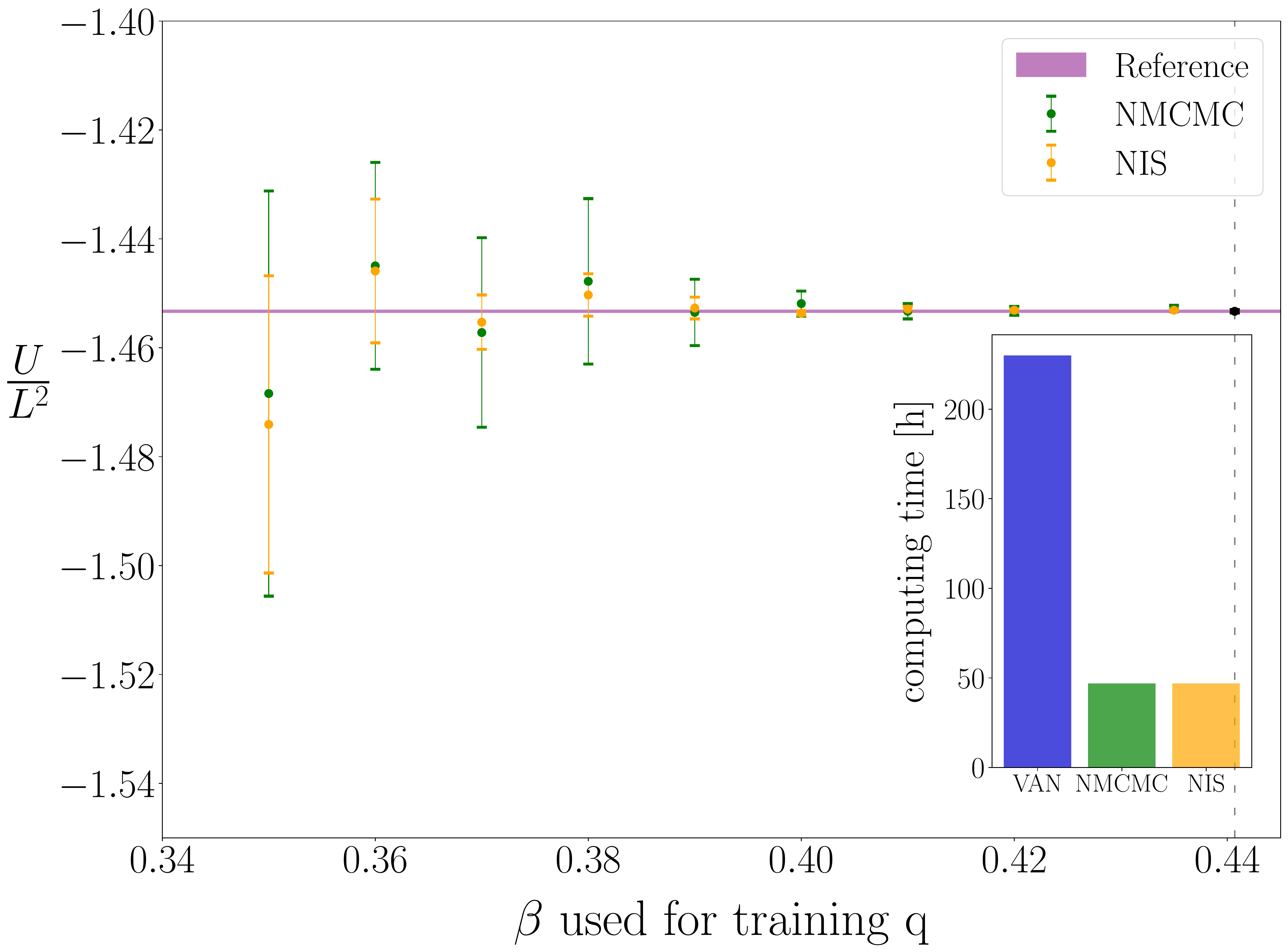}
    \caption{\label{fig:transfer} Samplers $q$ are trained at increasingly lower $\beta$ values and used to predict the internal energy $U/L^2$ at the critical coupling $\beta_c$. All results agree with the reference values within error bars. The variance of the estimators increase as the difference between model and target temperature gets larger. Transfer runs for NMCMC and NIS allow to only train one model which leads to significant speed up since runtime is dominated by training, as illustrated in the inset.}
\end{figure}

Lastly, our proposed methods allow for transfer across parameter space, as explained in Section~\ref{sec:methods}. In Figure~\ref{fig:transfer}, we summarize a few transfer runs. 
We performed a full training procedure for each value of $\beta$ shown in Figure 6. The trained samplers were then used to estimate observables at $\beta_c$ (i.e. \textit{not} at the training temperature $\beta$).
All predicted values agree with the reference within error bars. As the difference between model's temperature $\beta$ and target $\beta_c$ increases, the variance grows as well --- as was to be expected. In practice, this limits the difference between model and target inverse temperature. Nevertheless, we can use models trained at a single $\beta$ value to predict other values in a non-trivial neighbourhood of the model $\beta$. This allows to more finely probe parameter space at only minimal additional computational costs. 

\subsection{Neural MCMC}
NMCMC obtains a proposal configuration by independent and identically distributed sampling from the sampler $q$. This can result in a significantly reduced integrated autocorrelation time $\tau_{int, \mathcal{O}}$ for the observables $\langle \mathcal{O} \rangle$. For this reduction, it is not required to perfectly train the sampler. It is however required that the sampler is sufficiently well-trained such that the proposal configuration is accepted with relatively high probability, as illustrated in Figure~\ref{fig:autocorr}. Table~\ref{tab:autocorr} demonstrates a significant reduction in integrated autocorrelation $\tau_{int}$, as defined in \eqref{eq:autocor}, for two observables at $\beta_c$ on a $16 \times 16$ lattice. 

\begin{table}[ht]
  \begin{center}
    \caption{Neural MCMC instead of Metropolis leads to a significant reduction of integrated autocorrelation times $\tau_{int}$ for a $16 \times 16$ lattice at  $\beta_c$. The neural sampler was trained over ten thousands steps and the acceptance rate was 69 percent. The observables $U=\mathbb{E}_p[H]$ and $|M|=\sum_i\mathbb{E}_p(|s_i|)$ are the internal energy and the absolute magnetization respectively.}
    \label{tab:autocorr}
    \begin{tabular}{l|cc} 
    Observable & Metropolis & NMCMC \vspace{0.1cm}\\
    \colrule
    \textbf{$\tau_{int, U}$}  & 4.0415 & 0.8317 \\\vspace{0.05cm} 
    \textbf{$\tau_{int, |M|}$}  & 7.8510 & 1.3331 
    \end{tabular}
  \end{center}
\end{table}

In NMCMC, the proposal configuration $s \sim p_0(s|s') =q(s)$ is independent of the previous configuration $s'$ in the chain. This is in stark contrast to the Metropolis algorithm for which the proposal configuration is obtained by a local update of the previous configuration. As a result, NMCMC is less likely to stay confined in (the neighbourhood of) an energy minimum of the configuration space. This is demonstrated in Figure~\ref{fig:histogram} which shows the magnetization histograms for Metropolis and Neural MCMC. Since the Ising model has a discrete $\mathbb{Z}_2$-symmetry, we expect a two-moded distribution. In constrast to the Metropolis algorithm, NMCMC indeed shows such a behaviour.

\section{Applicability to Other Samplers}\label{sec:applic}
We note that our approach can in parts be applied to other generative models. Table~\ref{tab:taxonomy} summarizes the applicability of neural MCMC (NMCMC) sampling and neural importance sampling (NIS).
Namely, when the employed GNS provides an unnormalized sampling probability, i.e., the exact probability multiplied by a constant, then NMCMC and NIS can again correct the learned sampler $q$ leading to asymptotically unbiased estimators. However, the applicability is limited to the observables that do \emph{not} explicitly depend on the partition function, i.e., $h \equiv 0$ in Eq.~\eqref{eq:GeneeralObservable}.

If the employed GNS allows us to approximate the (normalized or unnormalized) sampling probability, one can apply our approach by using the approximate probability for $q$.
The bias can then be reduced if the gap between the target distribution and the sampling distribution is larger than the approximation error to the sampling probability.  However, then the estimator may not be asymptotically unbiased.

In summary, our method can be applied broadly to a variety of machine learning models and therefore does not depend on the particular choice of the sampler. Depending on the physical system, a particular architecture may be preferable. For example, whether an autoregressive model or a normalizing flow is the best machine learning tool to use, depends on how the sampling space looks like. The former requires a discrete sampling space, thus being a particularly good match for discrete systems, such as spin chains; the latter finds its applicability in the context of continuous systems, such as lattice field theories.
As shown in Table~\ref{tab:taxonomy}, applying our method to these models provides 
asymptotically unbiased estimators.

\section{\label{sec:conclusion}Conclusion and Outlook}
In this work, we presented a novel approach for the unbiased estimation of observables with well-defined variance estimates from generative neural samplers that provides the exact sampling probability (GNSEP). Most notably, this includes also observables that explicitly depend on the partition function such as free energy or entropy. The practical applicability of the approach is demonstrated for the two-dimensional Ising model, stressing the importance of unbiased estimators compared to biased estimators from the literature.

In summary, the methods proposed in this paper not only lead to theoretical guarantees but are also of great practical relevance. They are applicable for a large class of generative samplers, easy to implement, and often lead to a significant reduction in runtime. We therefore firmly believe that they will play a crucial role in the promising application of generative models to challenging problems of theoretical physics.

\begin{table*}[ht]
\centering  
    \caption{Applicability of NMCMC and NIS to various GNSs. $h$ refers to the term explicitly depending on the partition function $Z$ of the observable (see Eq.~\eqref{eq:GeneeralObservable}). General Adversarial Networks (GANs) do not provide sampling probabilites and therefore can not be used for our method. Restricted Boltzmann Machines (RBMs) only provide approximate and unnormalized sampling probability and therefore do not lead to asymptotically unbiased estimators using our methods. Because of the lack of normalization, observables with explicit dependence on the partition function cannot be estimated. Variational Autoencoders (VAEs) provide approximate sampling probability. Our method can therefore be applied but does not lead to asymptotic guarantees. The cases of Normalizing Flows (NFs) and Autoregressive Models (AMs) were discussed at length before. The applicability is summarized in the table using the following notation: $\checkmark$: the estimator is asymptotically unbiased;  (\checkmark): applicable but the estimator is still biased;  \text{\sffamily X}: not applicable. Generative neural samplers in the last row are GNSEP as introduced in Sec.~\ref{sec:VAN}. The last column gives example references in which the corresponding type of GNS is applied to a physical system.} 

    \label{tab:taxonomy}
    \begin{tabular}{c|c|c|c|c} 
    Accessible sampling probability & NMCMC, NIS($h \equiv 0$) & NIS($h \ne 0$) & GNSs & Application in Physics \\
    \hline
    none & \text{\sffamily X} & \text{\sffamily X} & GAN & \cite{zhou2019regressive,Urban:2018tqv} \\
        
    approximate, unnormalized &(\checkmark) & \text{\sffamily X} & RBM & \cite{carleo2017solving,Morningstar2017DeepLT} \\
     
    approximate, normalized & (\checkmark) &  (\checkmark) & VAE & \cite{Cristoforetti2017TowardsMP}\\
    
     exact, unnormalized & \checkmark & \text{\sffamily X} &  -- & -- \\ 

     exact, normalized & \checkmark & \checkmark & AM, NF & \cite{wu2019solving, noe2019boltzmann,albergo2019flow} \\ 

	\end{tabular}
\end{table*}

\acknowledgements
This work was supported by the German Ministry for Education and Research as Berlin Big Data Center (01IS18025A) and Berlin Center for Machine Learning (01IS18037I). This work is also supported by the Information \& Communications Technology Planning \& Evaluation (IITP) grant funded by the Korea government (No. 2017-0-001779) and by the DFG (EXC 2046/1, Project-ID 390685689). Part of this research was performed while one of the authors was visiting the Institute for Pure and Applied Mathematics (IPAM), which is supported by the National Science Foundation (Grant No. DMS-1440415). The authors would like to acknowledge valuable comments by Frank Noe and Alex Tkatchenko to an earlier version of the manuscript.

\appendix
\section{Partition function of the finite-size Ising model}\label{app:ZIsing}
In this appendix, we review the exact solution for the partition function of the finite-size Ising model \cite{ferdinand1969bounded}. For an $L \times L$ lattice, the partition function is given by
\begin{align}
    Z = \frac12 (2 \sinh (2 \beta))^{L^2/2} \sum_{i=i}^4 Z_i \,,
\end{align}
where we have used the definitions
\begin{align}
    &Z_1 = \prod_{r=0}^{L-1} 2 \cosh (\tfrac12 L \gamma_{2r+1}) \,, &&  Z_2 = \prod_{r=0}^{L-1} 2 \sinh (\tfrac12 L \gamma_{2r+1}) \,, \nonumber\\
    & Z_3 = \prod_{r=0}^{L-1} 2 \cosh (\tfrac12 L \gamma_{2r}) \,, &&  Z_4 = \prod_{r=0}^{L-1} 2 \sinh (\tfrac12 L \gamma_{2r}) \,,
\end{align}
with the coefficients
\begin{align}
&\gamma_0 = 2 \beta + \ln \tanh \beta \,, \nonumber\\
&\gamma_r = \ln (c_r + \sqrt{c_r^2-1}) && \text{for} && r>0 \,,
\end{align}
and $c_r = \cosh 2\beta \coth 2 \beta - \cos(r \pi/L)$. From this expression for the partition function, one can easily obtain analytical expressions for the free energy and entropy.

\section{Proof of asymptotic Unbiasedness}\label{app:proofs}
In this section, we will give a review of the relevant arguments establishing that the NIS and NMCMC estimators are asymptotically unbiased.

For reasons that will become obvious soon, it is advantageous to re-interpret the original network output $q' \in [0, 1]$ as the probability $q \in [\epsilon, 1-\epsilon]$ by the following mapping:
\begin{equation}
q = \left(q' - \frac{1}{2}\right) \left(1 - 2 \epsilon\right) + \frac{1}{2}.
\end{equation}

Due to the rescaling discussed above, we can assume that the support of the sampling distribution $q$ contains the support of the target distribution $p$. This property is ensured since the sampler $q$ takes values in $q \in [\epsilon, 1-\epsilon]$.
\subsection{Neural Importance Sampling}
Importance sampling with respect to $q$, i.e. 
\begin{align}
    &\mathbb{E}_p[ \mathcal{O}(s) ] \approx \sum_{i=1}^N w_i \, \mathcal{O}(s_i) \,,\, \text{where}\nonumber\\
    &s_i \sim q(s) \,, \, w_i = \frac{\hat{w}_i}{\sum_i \hat{w}_i} \,, \, \hat{w}_i = \frac{e^{-\beta H(s_i)}}{q(s_i)} \,,
\end{align}
is an asymptotically unbiased estimator of the expectation value $\langle \mathcal{O}(s) \rangle$ because
\begin{align}
    \mathbb{E}_p[ \mathcal{O}(s) ] &= \sum_s p(s) \mathcal{O}(s) = \sum_s q(s) \, \frac{p(s)}{q(s)} \,  \mathcal{O}(s)\nonumber\\ 
    &= \frac{1}{Z}  \sum_s q(s)\, \underbrace{\frac{\exp(-\beta H(s))}{q(s)}}_{=\hat{w}(s)} \,  \mathcal{O}(s)\nonumber \\
    &= \frac{1}{Z N}  \sum_{i=1}^N \hat{w}(s_i) \mathcal{O}(s_i) + o_P(1)  \,,
\end{align}
where $s_i \sim q$. The partition function $Z$ can be similarly determined
\begin{align}
    Z &= \sum_s \exp(-\beta H(s))\nonumber\\ 
    &= \sum_s q(s) \frac{\exp(-\beta H(s))}{q(s)} = \frac{1}{N} \sum_{i=1}^N  \hat{w}(s_i) + o_P(1)\,.
\end{align}
Combining the previous equations, we obtain 
\begin{align}
 \langle \mathcal{O}(s) \rangle_{p} = \sum_{i=1}^N w_i \, \mathcal{O}(s_i) + o_P(1) && \text{with} && w_i = \frac{\hat{w}_i}{\sum_i \hat{w}_i} \,.
\end{align}

\subsection{Neural MCMC}
The sampler $q$ can be used as a trial distribution $ p_0(s' | s) = q(s')$ for a Markov-Chain which uses the following acceptance probability in its Metropolis step
\begin{align}
    p_a(s'| s) &= \text{min}\left(1, \frac{p_0(s| s') p(s')}{p_0(s'| s) p(s)}\right)\nonumber\\  
    &= \text{min}\left(1, \frac{q(s) \, \exp(-\beta H(s'))}{q(s') \, \exp(-\beta H(s))}\right)  \,. 
\end{align}
This fulfills the detailed balance condition
\begin{align}
    p_t(s'|s) \exp(-\beta H(s)) = p_t(s|s') \exp(-\beta H(s))
\end{align}
because the total transition probability is given by $p_t(s'|s)=q(s') \, p_a(s'|s)$ and therefore
\begin{align}
     p_t&(s'|s) \exp(-\beta H(s)) \nonumber\\
     &= q(s') \,  \text{min}\left(1, \frac{q(s) \, \exp(-\beta H(s'))}{q(s') \, \exp(-\beta H(s))}\right) \,  \exp(-\beta H(s)) \nonumber\\ 
     &= \text{min}\left\{q(s') \exp(-\beta H(s)), \, q(s)\exp(-\beta H(s'))\right\} \nonumber\\
     &= p_t(s|s') \exp(-\beta H(s')) \,,
\end{align}
where we have used the fact that the min operator is symmetric and that all factors are strictly positive. The latter property is ensured by the fact that $q(s) \in [\epsilon, 1-\epsilon]$. 

\section{Variance Estimators}\label{app:variance}
\begin{table}[ht]
  \begin{center}
    \caption{Comparison of standard deviation estimated as in Section~\ref{sec:VarianceEstimators} to sample standard deviation over ten runs. Experiments are done at an inverse temperature $\beta=0.44$.
    }
    \label{tab:varEstimators}
    \begin{tabular}{l|cc} 
    Observable & 
    Estimated Std & Sample Std \vspace{0.1cm}\\
    \colrule
    \textbf{Entropy}   & 0.00023 & 0.00025 \\

    \textbf{Free En.}   & 0.00002 & 0.00002 
    \end{tabular}
  \end{center}
\end{table}
As explained in the main text, we estimate observables of the form
\begin{equation}
    \mathcal{O} = g(s) + h(Z)
\end{equation}
by the samples $s_i \sim q$ with $i=1\dots N$ using
\begin{align}
    \hat{\mathcal{O}}_N = \frac{\tfrac{1}{N}\sum_{i=1}^N \mathcal{O}(s_i, \hat{Z}_N) \, \hat{w}(s_i)}{\hat{Z}_N} \,.
    \label{eq:NIPEstimator}
\end{align}
By the definition of $\hat{Z}_N$, see \eqref{eq:estimatorZ}, this is equivalent to
\begin{equation}
    \hat{\mathcal{O}}_N =
\frac{   \frac{1}{N} \sum_{i=1}^N g(s_i) \hat{w}(s_i) }{\hat{Z}_N} + h( \hat{Z}_N).
\end{equation}

Let
\begin{align}
\boldsymbol{\phi}_N
&=
\frac{1}{N} \sum_{i=1}^N
\boldsymbol{\phi}(s_i)
\qquad
\mbox{for}
\qquad
\boldsymbol{\phi}(s) =
\begin{pmatrix}
 g(s) \hat{w}(s)  \\
 \hat{w}(s)
 \end{pmatrix}.
\end{align}
Then, 
the central limit theorem implies that
\begin{align}
\boldsymbol{\phi}_N 
 \xrightarrow[]{D}
 \mathcal{N} \left(
 \boldsymbol{\phi}^*,
 \frac{1}{N} \boldsymbol{\Sigma} \right),
 \end{align}
 where
 \begin{align}
 \boldsymbol{\phi}^* &=  \mathbb{E}_q \left[\boldsymbol{\phi}\right]
 = 
 \begin{pmatrix}
\mathbb{E}_p[g] \, Z\\
Z
 \end{pmatrix}
 ,
 \notag\\
 \boldsymbol{\Sigma}
 &= \mathbb{E}_q \left[
\boldsymbol{\phi} \boldsymbol{\phi}^T
 \right]
 -  \mathbb{E}_q[\boldsymbol{\phi}] \mathbb{E}_q[\boldsymbol{\phi}^{T}].
\end{align}

Since the estimator \eqref{eq:NIPEstimator}
can be written as a smooth function $f$
of $\boldsymbol{\phi}_N$ as
\begin{align}
f(\boldsymbol{\phi}_N) := \frac{(\boldsymbol{\phi}_N)_1}{(\boldsymbol{\phi}_N)_2} + h\big((\boldsymbol{\phi}_N)_2\big) = \hat{\mathcal{O}}_N,
\end{align}
its variance can be written as
\begin{align}
\sigma^2_{\hat{\mathcal{O}}_N}
&= 
\frac{1}{N}
\boldsymbol{\psi}^T \,
\boldsymbol{\Sigma}^{\vphantom{T}} \,
\boldsymbol{\psi}^{\vphantom{T}}
+ o_P(N^{-1})
\label{eq:VarianceFirst}
\end{align}
with 
\begin{align}
    \boldsymbol{\psi} &= \nabla f (\boldsymbol{\phi}^*) = \left. \begin{pmatrix}
1 / (\boldsymbol{\phi}_N)_2 \\
- (\boldsymbol{\phi}_N)_1 / (\boldsymbol{\phi}_N)_2^2
+ h' \big((\boldsymbol{\phi}_N)_2 \big) 
 \end{pmatrix} \right|_{\boldsymbol{\phi}^*}\nonumber\\
 &= \begin{pmatrix}
1 / Z \\
- \mathbb{E}_p[g] / Z
+ h' \left(Z \right)
 \end{pmatrix}.
\end{align}
In Table~\ref{tab:varEstimators}, we numerically verify that our estimated standard deviation is consistent with the sample standard deviation over ten runs.

\section{Experimental Details}\label{appendix:hyperparams}
In this appendix, we provide an overview of the setup used for the experiments presented in this manuscript. 
\subsection{Model Training}
Unless reported differently, all the models were trained for a $16\times16$ lattice for a total of $10000$ steps. The model trained on a $24 \times 24$ lattice required $15000$ steps until convergence.
Our model use the VAN architecture with residual connections (see \cite{wu2019solving} for details on this architecture). The networks are six convolutional layers deep (with a half-kernel size of three) and has a width size of 64. A batch size of $2000$ and a learning rate of $10^{-4}$ were chosen. No learning rate schedulers were deployed in our training. For each model, we applied \textit{$\beta$-annealing} to the \textit{target} $\beta_t$ using the following annealing schedule
\begin{equation}
\beta = \beta_t (1-0.998^{N_{step}})
\end{equation}
where $N_{step}$ is the total number of training steps. We summarize the used setup in Table~\ref{tab:hyperparameters}. Training a sampler for a $16 \times 16$ and $24 \times 24$ lattices takes respectively $\sim17$ and $\sim75$ hours of computing time on three Tesla P100 GPUs with 16GB VRAM. As specified before, the former lattice required ten thousands steps of training to converge whereas the latter required fifteen thousands.\\

\begin{table}[ht]
  \begin{center}
    \caption{Summary of the hyperparameters setup used for training our samplers.} \label{tab:hyperparameters}
    \begin{tabular}{cccccccc} 
    Sampler & Depth & Width & Batch & lr & Steps & $\beta$ Ann.\\
    \colrule
    PixelCNN  & 6 & 64 & $2\cdot 10^{3}$ & $10^{-4}$ & $10^{4}$ & 0.998 \\
    \end{tabular}
  \end{center}
\end{table}

\begin{table}[ht]
  \begin{center}
    \caption{Internal energy per lattice site on an 8$\times$8 lattice. Values which overestimate the ground truth $U/L^2=-1.54439$ are in bold. Same holds for Entropy, ground truth $S/L^2=0.25898$. The second row shows absolute magnetization with ground truth $|M|/L^2=0.8083$. In this case, estimates which overestimate the ground truth are in bold.}
    \label{tab:5VANat0.45}
    \begin{tabular}{l|ccccc} 
    Obs & Model 1 & Model 2 & Model 3 & Model 4 & Model 5\\
    \colrule
    $U / L^2$ & \textbf{-1.5407} & -1.5461 & \textbf{-1.5364} & \textbf{-1.5438} & \textbf{-1.5421}\\
    $|M| / L^2$ & \textbf{0.8089} & \textbf{0.8114} & 0.8059 & \textbf{0.8098} & 0.8070\\
    $S / L^2$ & 0.258889 & \textbf{0.260271} & 0.257843 & \textbf{0.262209} & \textbf{0.261478} \\
    \end{tabular}
  \end{center}
\end{table}

\subsection{Neural Monte Carlo and Neural Important Sampling}
In Neural MCMC, we use a chain of $500$k steps. Conversely to standard MCMC methods, such as Metropolis, no equilibrium steps are required since we sample from an already trained proposal distribution. In Neural Importance Sampling, batches of 1000 samples are drawn 500 times. Both sampling methods were performed on a Tesla P100 GPU and their runtime is approximately an hour in the case of a $16 \times 16$.

\section{Direction of Bias}\label{app:directionofbias}

In this appendix, we demonstrate that the direction of the bias depends on the random initialization of the network. In order to illustrate this fact, we trained five models at $\beta=0.45$ for a 8$\times$8 lattice using the same hyperparameter setup. We compare the estimate of the energy with an exact reference value of $U/L^2=1.54439$. Table~\ref{tab:5VANat0.45} summarizes the results. Values which overestimate the ground truth are in bold. This shows that the trend of under- or overestimating, observed from Figure~\ref{fig:observables}, holds only on average and is not a systematic effect.\\

\newpage
\bibliographystyle{apsrev4-1}
\bibliography{literature}

\end{document}